**Center of mass perturbation as a normalizable estimate of dynamic balance in gait: an application comparing typically developing children with spastic cerebral palsy**

*Submitted as an Original Article by*

Timothy Niiler[§], John Henley, Freeman Miller

[§]Corresponding author

Gait Laboratory, Nemours/Alfred I. duPont Hospital for Children,

P.O. Box 269, Wilmington, DE 19899

Phone: 302-651-4615, Fax: 302-651-5144

Email Addresses:

  Timothy Niiler, PhD – Corresponding Author - tim.niiler@gmail.com

  John Henley, PhD – john.henley@nemours.org

  Freeman Miller, MD – freeman.miller@nemours.org

**ABSTRACT**

It is well recognized that the relative position of the center of mass (pCOM) with respect to the base of support (BOS) is a determining factor in the maintenance of balance. However, during gait the dynamic nature of the BOS is not well defined and in most studies is completely ignored. Prior work tends to focus on the variability in the position of the center of mass (COM) with respect to the laboratory reference frame to attempt to quantify dynamic balance. We propose a modified method whereby the position of the COM in an end-effector based coordinate system may be used as an improved estimate of threat to balance during gait. The distance ($D_N$) of the projection of the COM (pCOM) from the reference axis, normalized by half the foot length for inter-subject comparison, is an estimate of the gravitational moment arm. It can be shown that in concordance with theory, balance impaired subjects with spastic cerebral palsy (CP) have a larger $D_N$ than their typically developing peers. Furthermore, when compared with variational methods, $D_N$ shows better discriminative ability in characterizing groups.

**1. Introduction**

A number of methods that have been proposed to quantify dynamic balance in gait. These methods include tracking the variability of the center of mass (COM) trajectory [1,2,3,4], calculating a margin of stability [5,6], using non-linear dynamics such as Lyapunov exponents or Floquet multipliers to compare stability [7,8], and tracking standing balance with the presumption that it correlates with dynamic balance [9]. All of these methods are based, at least in part, on COM motion with respect to some fixed reference. This reference can be the laboratory frame [7], the center of pressure [9,10,11], the edge of the grounded base of support [5], or an initial COM position [10,11]. When variability is studied, the reference is often the range within a gait cycle.

Studies have shown that individuals with balance impairments have larger medio-lateral variability in their COM than do healthy controls [3,10,11,12,13]. However, this is only a relative measure and provides little information on how balanced or unbalanced the healthy controls actually are. Ultimately, one cannot use COM as an indicator of balance outside the context of the base of support. In typical balanced gait, the projection of the COM (pCOM) tracks between the feet at all points of the gait cycle moving from the inside edge of the support foot during single support, traversing towards the other foot at heel strike and then repeating this cycle with the contralateral foot [14]. While the same essential trajectory of the pCOM is followed in CP gait [10,11], there are deviations which relate to differences in foot placement and posture. This implies that such perturbations may be a proxy of threat to balance. Therefore, to properly quantify deviations in pCOM trajectory, one must first establish a coordinate system that reflects the positioning of the pCOM with respect to the feet.

The use of the feet as a reference frame, or equivalently the base of support (BOS), is regularly done in static balance where, on average, the COP and the pCOM are found to be between the feet halfway between heel and toe [5]. As such, why not use the line joining the centers of the feet, what we will call the inter-foot line (IFL), as the basis for reference with regards to dynamic balance? In this paper we will introduce $D_N$ which is a measure (an index) based on this dynamic reference and show how it can be applied. We will demonstrate the validity of $D_N$ by comparing typically developing individuals to individuals whose balance is impaired due to cerebral palsy (CP). We postulate that for typical balanced gait, pCOM tracks close to the IFL making the IFL a reasonable reference frame for the measurement of balance. Similarly, we postulate that in for those whose gait is altered by CP, $D_N$ will be significantly larger than for typically developing controls. Finally, we hypothesize that the variability in $D_N$ will also be larger for the CP population compared to the TD controls.

**2. Methods**

*2.1 Dynamic Balance Index $D_N$*

As we are suggesting that balance can be characterized by the relationship between the base of support (BOS) and the pCOM, we need to characterize the dynamic base of support that is present in walking. We start with defining the center as the inter-foot-line (IFL) as the line that passes through the bisection of heel and toe markers of each foot. Then the relationship between the pCOM and the base of support can be characterized by $D_{IFL}$ the shortest distance between the pCOM and the IFL.

To account for differences in the size of individuals and their base of support, we normalize $D_{IFL}$ by half the foot length. If we let

1) $$D_N = \frac{D_{IFL}}{L/2}$$

where L is the average length of the foot, then when the magnitude of $D_N$ exceeds unity, the pCOM lies outside of the projection of the region between the feet. Note that our definition does not require that both feet be on the ground at all times. During the time when a foot is not on the ground the region under it is essentially part of the BOS as the foot can almost instantaneously be placed on the ground. We believe that this better characterizes the dynamics of the *functional* BOS during walking than simply defining the BOS to be only the region actual foot ground contact. When this happens and both feet are grounded in an attempt to recover balance, $D_N$ represents a normalized gravitational moment arm. Note that $D_N$ is positive if the moment arm lies forward of the IFL and negative if the moment arm is behind the IFL.

*2.2 Subjects*

After obtaining IRB approval, retrospective kinematic data was pulled for 14 subjects (aged 10.2 ± 5.1 years) with hemiplegic (n=13) and diplegic (n=1) CP prior to having single event multilevel surgery. All data were collected with an 8 camera motion capture system (Motion Analysis Corporation, Santa Rosa, CA) at 60 Hz. Each person had a modified Cleveland Clinic marker set. Our choice of comparison group was governed by the CDC definition of CP as affecting balance [15], and evidence of balance deficits in this population [10,11,16,17]. All CP data was for a self-selected normal walking speed. For reference, retrospective kinematic data was pulled from the 66 typically developing (TD) subjects (aged 9.2 ± 4.3 years). In this case, we had data for self-selected slow (41 subjects), normal (43 subjects), and fast walking speeds (45 subjects). Not all of the TD subjects had data collect for each condition.

*2.3 Analysis*

We calculated the average $D_N$ at each point in the gait cycle for each subject and condition (TD: slow, normal, or fast speeds; CP: normal speed) in the analysis. Figure 2 shows a representative $D_N$ curve for a TD subject. Side to side asymmetry similar to that shown here was present in nearly every subject regardless of condition. For this reason, to avoid washing out the magnitude of side to side changes in $D_N$, curves were lined up so that the side with the minimum $D_N$ value at heel strike came first prior to averaging curves between subjects. Data were modeled using locally weighted regression smoothing (LOESS), and from this model representative mean and standard deviation values were computed on a point-wise basis per group. Modeled mean $D_N$ curves were then compared between CP and typically developing conditions using a Welch's two-sample t-test (for unequal variances) at each point within the gait cycle to indicate where in the gait cycle the conditions were the most similar. Due to LOESS, p-values of adjacent testing were highly correlated and use of Bonferroni corrections, which assume independence in testing, were not used except between groups.

## 3. Results

Average velocities for each group were 86.3 ± 20.8 cm/s for the slow TD group, 115.6 ± 21.3 cm/s for the normal TD group, 170.1 ± 26.3 cm/s for the fast TD group, and 68.1± 26.1 cm/s for the CP group. Figure 2 shows a coronal plane view of the progression of the $D_{IFL}$ during gait for representative typically developing and CP subjects. A couple of things are immediately noteworthy: the stance of the TD subject is narrower than that of the CP subject, and the $D_N$ of the CP subject is considerably larger at all phases of the gait cycle. This has consequences for the pCOM which tracks further out from the interior boundaries of the feet for the CP subject than for the TD subject.

The samples shown in Figure 2 are consistent with the graph in Figure 3a where average $D_N$ is substantially larger throughout the gait cycle for the CP group than for any of the TD groups. The $D_N$ curves are all double-humped, regardless of population, with heel strikes corresponding to the troughs in the graph and swing leg cross-over corresponding to the peaks. The graphs are asymmetrical with heel-strike about the $D_N = 0$ line. Very few subjects were symmetrical between left and right sides in their average $D_N$ graphs. Also, the CP subjects' pCOM was generally ahead of the IFL compared to the TD subjects whose pCOM oscillated about this line. Figure 3a indicates that for TD subjects, the pCOM remains very close to the IFL on average, but that it moves further forward with respect to this line as velocity increases. Average $D_N$ was -0.157 ± 0.118 for the slow TD group (avg range [0.190, 0.099]), 0.002 ± 0.120 for the normal TD group (avg range [-0.194, 0.151]), and 0.084 ± 0.151 for the fast TD group (avg range [-0.173, 0.319]), and 0.538 ± 0.243 for the CP group (avg range [0.039, 0.929]). The key increase in $D_N$ with speed occurs at swing leg cross-over as indicated by the increase in peaks in Figure 3a.

Figures 3b and 3c present the results of per-condition t-tests at each point in the gait cycle to highlight where differences in groups were *most* significant. These results are presented using a natural log of the p-value since the p-values were for the most part very much less than 0.05, the standard for significant differences. On such a graph, for a single comparison with α equal to 0.05, values less than -2.99 ( =ln(0.05) ) are significant. In our case, since there were multiple comparisons, α is obtained from the conservative Bonferroni correction (α = 0.05/3 comparisons), so values less than -4.09 are significant. As per Figure 3b, all TD groups were significantly different from the CP group at all points in the gait cycle with the fast speed group being the most similar and the slow speed group being the most different. Figure 3c indicates that the TD groups tended to be significantly different from each other primarily during swing phase of gait. As might be expected, the fast speed group was most different from the slow speed group.

**4. Discussion**

Prior study of dynamic balance in individuals with CP compared to TD controls has shown that the variability of the relative positioning of the COM is greater in the CP group [10]. The current study reinforces such results in the motion of the pCOM with respect to the IFL. The trend holds for both range and standard deviations of the data at any point in the gait cycle (Figure 4). Furthermore, the parameter $D_N$ shows better discrimination between groups than does the typical analysis of COM displacement. Table 1 presents data from Hsue et al. [10] in comparison to data from this study. While there is a clear difference between Hsue et al.'s diplegic CP group and their TD group based on normalized COM displacement ($\Delta$COM) in the medio-lateral and vertical directions, this change is much more noticeable with $D_N$. In fact, there is a factor of two difference between average maximum $D_N$ values for fast versus normal speeds and normal versus slow speeds. If this sensitivity to condition holds for other groups, it may be that $D_N$ is clinically more useful than the medio-lateral/vertical displacements of the COM in the detection of changes in balance due to interventions or differences in balances between groups.

It was also anticipated that $D_N$ would be significantly larger for the CP group compared to the TD controls since the CP population is known to have impaired balance. Subjects with CP also typically have some degree of crouch, a condition which puts the pCOM forward of the IFL. This, in turn, causes the deviations from the typical path of the COM which we take to be indicative of increased threat to balance compared to the TD controls. When an individual with CP drags a toe and subsequently trips, the larger $D_N$ implies that there is both a larger gravitational moment acting to unbalance them. Likewise, there is a smaller margin for recovery since the pCOM will pass beyond the base of support sooner.

While $D_N$ is seen to increase with walking speed, the increase occurs primarily at mid-stance when $D_N$ is pointed in the direction of motion. This implies that the COM positioning at mid-swing is what is used to increase velocity as noted by Winter [14]. Yet our data also support previous findings that the pCOM remains between the feet during the entire gait cycle regardless of speed. Therefore, speed does not specifically predict instability since as the pCOM moves faster, it still maintains a similar relationship to the foot positioning. There are correlations between static balance and velocity in the stroke and diabetic populations: individuals with better balance tend to walk faster [18,19]. But if this is true, then explicit consideration of velocity may be misleading when one considers threat to balance. It is not just the velocity of the COM that threatens balance, but the relative positioning of the COM with respect to the BOS that matters. If one is able to control the relationship between the end effectors and the COM, then balance can be maintained. Loss of balance only ensues when one cannot control this relationship. For example, in a trip or slip, one loses the ability to control the base of support and hence the balance relationship. Here velocity determines how fast the relationship is distorted, and the momentum of the COM determines the impulse needed to restore balance. Someone who is moving faster when they trip will indeed have less time employ a recovery strategy before they fall.

Beyond empirical observation that for balanced gait the pCOM lies between the feet at all times and that the magnitude of $D_N$ is generally less than one, there are physical reasons that support the use of $D_N$ as a balance metric. Consider, for example, tripping, which is a leading cause of falls. There are two coping mechanisms for tripping: raising and lowering [20,21]. The raising strategy is used during early swing phase to try to clear the trailing foot. If successful, after a brief stutter, gait will continue as before. But if not successful, the feet are momentarily anchored to the ground while the COM will continue to move. If the body is treated as a rigid body subject to the constraints of the feet, the only degree of freedom for rotation is about the IFL. While the body may truly have more degrees of freedom, this assumption is consistent with the dynamic balance formulation of Hof et. al. [5] who use an inverted pendulum as a starting point and Buczek et. al. [22] who found inverted pendulum models to predict various kinematic and kinetic parameters of gait reasonably well. At this point during the trip, the gravitational COM moment arm can then be approximated by $D_{IFL}$.  On the other hand, when tripping occurs late in swing phase, the lowering strategy is used to bring the lead foot down as quickly as possible.  Once again, as the lead foot meets the ground the above constraints apply, and the relevant gravitational moment arm can be approximated by $D_{IFL}$.  The point that the relative positioning between end-effector and center-of-mass is fundamental to maintenance of balance is supported by Pijnappels et al. [21].  In their study of falling in younger and older subjects, they demonstrated that falling only occurred when the end-effectors were far behind the COM.

Another common mechanism for falling is slipping.  In this case the horizontal center of mass motion remains constant while the feet move out from under the COM.  Except in catastrophic slips, both feet are planted in an attempt to regain balance [1] and the relevant moment arm will again be $D_{IFL}$ as in tripping.  Now it may be that one or both feet come off the ground.  As the lead foot comes off the ground, a principle axis of the body's moment of inertia ellipsoid (MOIE) approximately tracks the orientation of the IFL due to the large mass of the legs.   An example illustrating the MOIE motion with respect to the body during gait may be seen in the HOAP2 simulation from Lee and Goswami [22].  As the fall continues and the second foot begins to leave the ground, the vector from the IFL to the pCOM will far exceed the boundary of the feet and it should be clear that the individual is unstable.

Perhaps surprisingly, $D_N$ was asymmetrical across the gait cycle for the TD subjects at all speeds.  While such asymmetry might be expected in the CP population due to differences in bilateral stiffness and motor control, these effects are not so substantial in the TD population.  Rather, the observed asymmetry may be the result of differences in limb-segment volume due to laterality, minor limb-length discrepancies, or differences in neurological control from side to side.  In any event, the asymmetry of $D_N$ in TD subjects merits further study.

It should be noted that $D_N$ is an approximation for the rough characterization of balance just as the gait deviation index (GDI) is a rough characterization of gait in general [23]. It does not attempt to quantify a subject's ability to exert muscle forces nor does it account for stiffness, spasticity, or even speed of fall. The margin of stability also the same limitations as well as several admitted approximations in its calculation: the assumption of a non-telescoping pendulum model, constants used for calculating the length of the pendulum are direction specific (sagittal plane vs frontal plane), and the inability to anticipate compensatory motions of the upper body [5]. Rather both $D_N$ and the margin of stability estimate threat to balance *in the moment* based on very simple assumptions. Unlike the margin of stability methods, $D_N$ has the advantage of being a normalized value which can, therefore, be compared between subjects. Compared to analysis of the variability of the COM, $D_N$ is more sensitive in showing differences between groups. Finally, $D_N$ can be computed from standard kinematic gait analysis data making it possible to use for dynamic balance assessment even when force plate data is not available. We have not yet validated $D_N$ with respect to the common, but more subjective, Berg Balance Scale and Dynamic Gait Index, but this should be considered as an important next step in assessing the clinical utility of this index.

*Table 1: Comparison of ΔCOM and maximum COM deviation results between TD and diplegic groups from Hsue et al [10] and this paper. All COM related values have been normalized to leg length for comparison and ΔCOM values were calculated as the difference between maximum and minimum average values. In Hsue et al, COM values were calculated with respect to the minimum center of pressure in a gait cycle, while in this paper, since there was no force plate data, COM values were calculated with respect to the mean COM per gait cycle. Both ΔCOM and maximum COM values are used since they are different estimates of balance. ΔCOM indicates variability throughout a gait cycle, while maximum values of COM deviation indicate points of maximum instability. Although maximum values of COM between Hsue et al. and this paper are not directly comparable due to a difference in reference point, the percent differences are generally comparable as the reference point represents a DC offset. Hsue et. al. values have been determined from their Figure 1 using custom graphical capture software. Differences in maximum to minimum $D_N$ results are also presented for this paper to show the increased sensitivity of this measure compared to COM excursion.*

|  | CP | TD | %Diff |
|---|---|---|---|
| Hsue et al (ΔCOM ML) | 0.0570 | 0.0310 | 59.1% |
| Hsue et al (ΔCOM Vertical) | 0.0410 | 0.0309 | 27.7% |
| Hsue et al (Max COM ML) | 0.173 | 0.103 | 50.7% |
| Hsue et al (Max COM Vertical) | 0.982 | 0.962 | 2.05% |
| This Paper (ΔCOM ML) | 0.0210 | 0.0114 | 59.2% |
| This Paper (ΔCOM Vertical) | 0.0494 | 0.0473 | 4.34% |
| This Paper (Max COM ML) | 0.0101 | 0.0057 | 55.7% |
| This Paper (Max COM Vertical) | 0.0216 | 0.0215 | 0.46% |
| This Paper (Δ$D_N$) | 0.929 | 0.151 | 87.8% |
| This Paper (Max $D_N$) | 0.929 | 0.151 | 144% |

Figure 1: Geometry illustrating the inter-foot-line and how $D_{IFL}$ is measured. This distance is the perpendicular along the ground from the IFL to the pCOM at any point during gait.

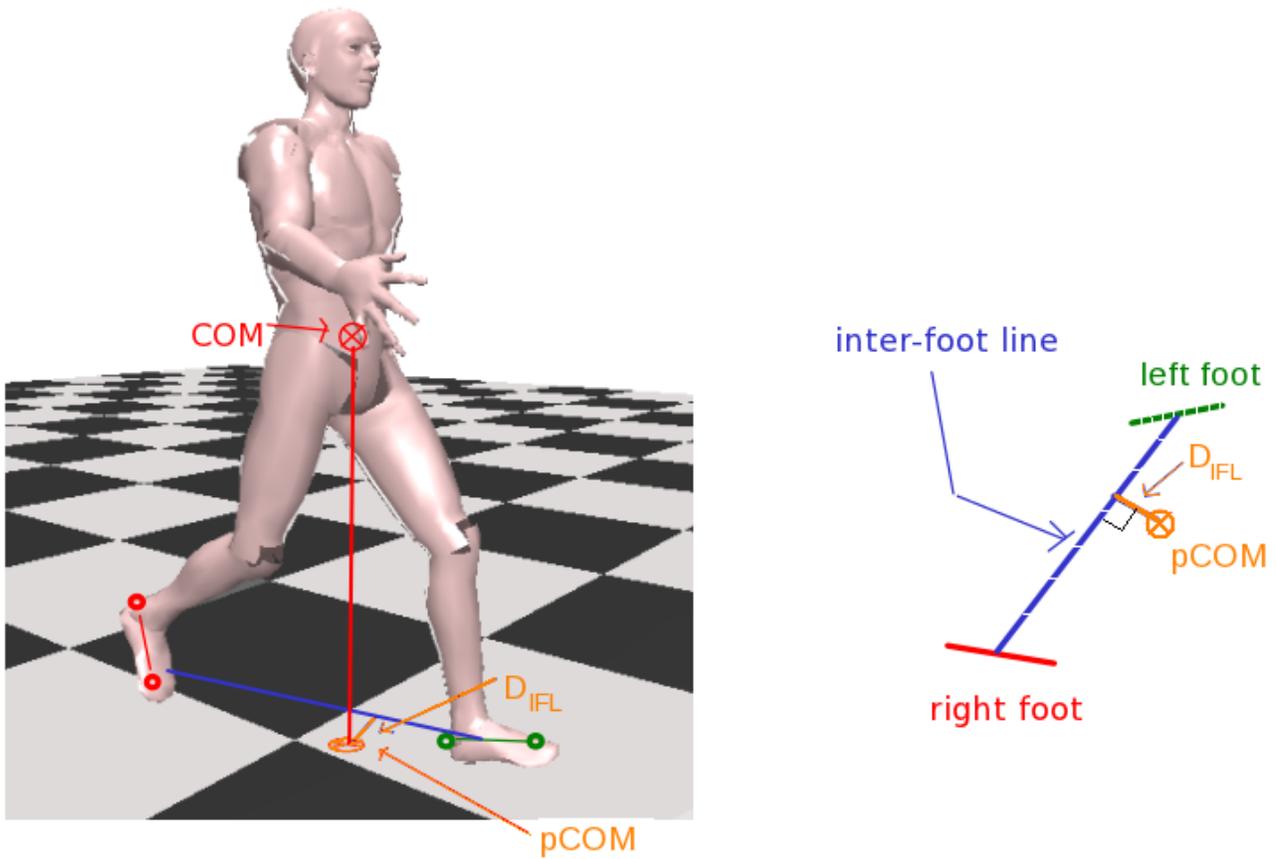

Figure 2: Transverse plane visualization of $D_{IFL}$ for representative typically developing (TD) and CP subjects showing a complete right and left side gait cycle (two right and left side heel strikes). Represented frames are equally spaced in time per subject. The timing difference between captured frames was selected to best show a range of positions without becoming too cluttered. The figure has been scaled horizontally for better comparison between conditions, and the direction of forward progression is to the right. Notice that $D_{IFL}$ is proportionally longer for CP gait than for TD gait at nearly every frame.

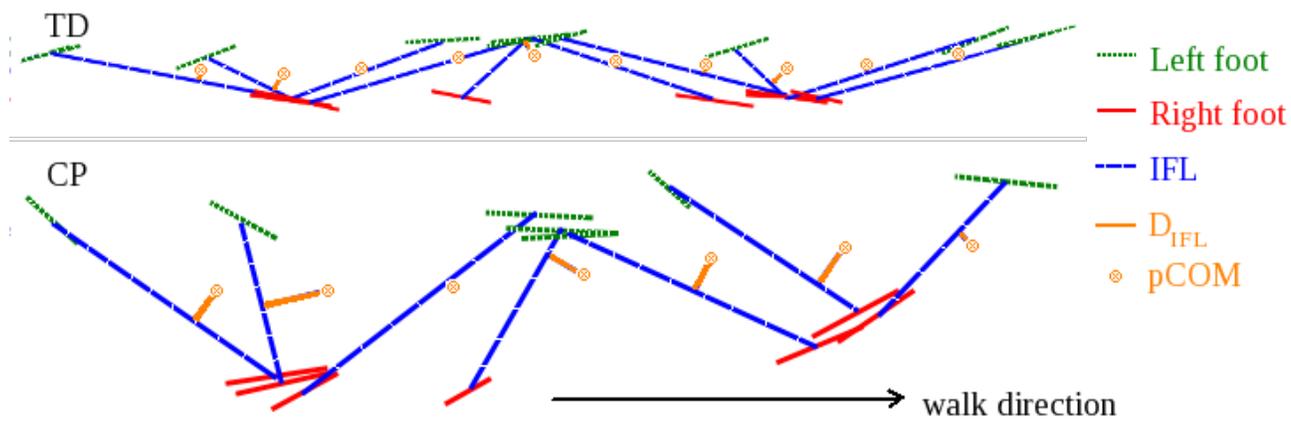

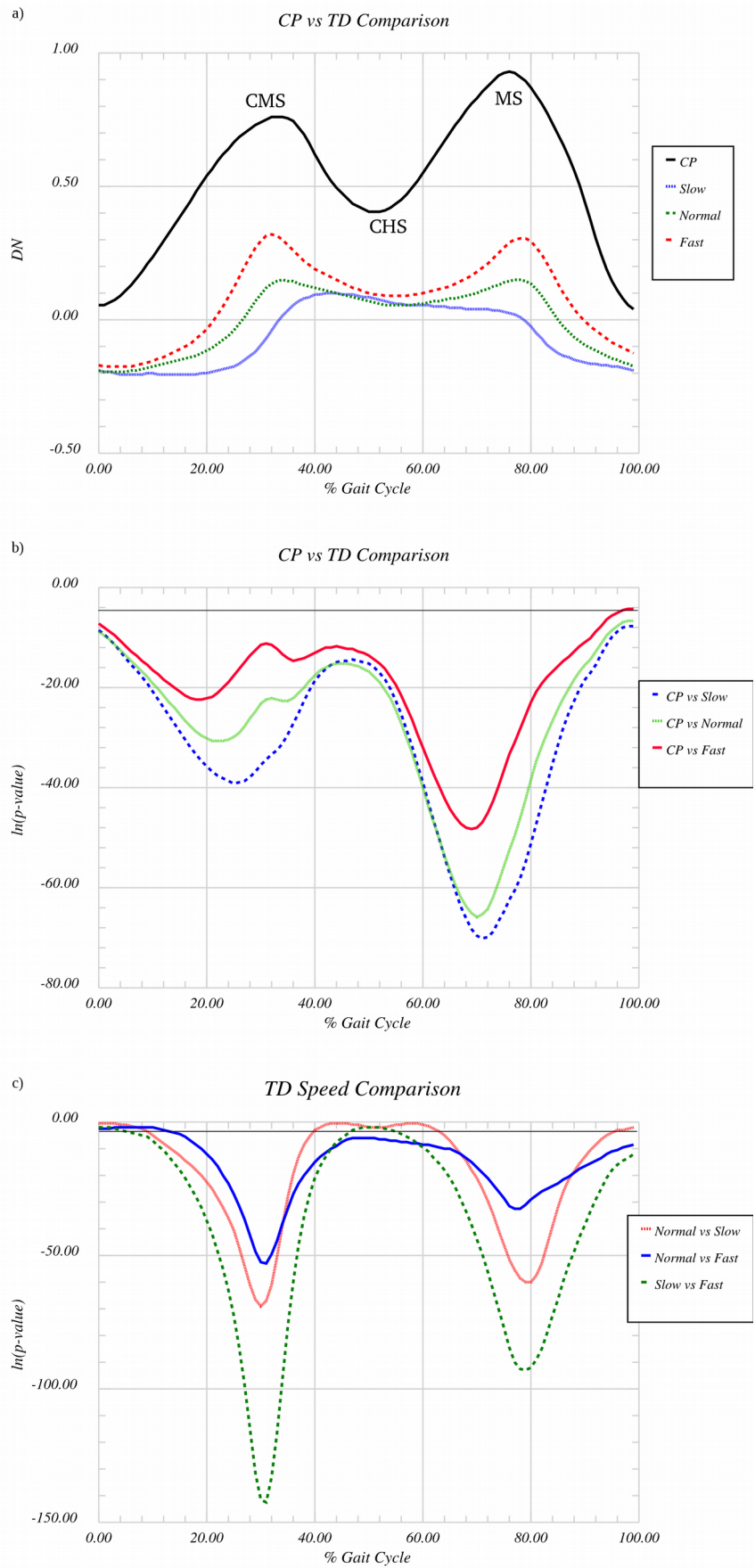

Figure 3: Results by condition. Figure 3a) shows the average $D_N$ across each condition from heel strike to heel strike. Additional labeled gait events are: Contralateral Mid Swing (CMS), Contralateral Heel Strike (CHS), and Mid Swing (MS). Figure 3b) shows natural log of p-values between CP and TD conditions as a function of gait cycle. Figure 3c) shows the natural log of p-values as a function of gait cycle in between TD conditions. The black horizontal lines in Figures 3b) and 3c) show a log(p-value) of -4.09 which corresponds to a Bonferroni corrected alpha level of 0.05/3. Values below this are significantly different.

Figure 4: Average ± 1 standard deviation of CP and TD normal speed subjects.  Despite similar trends, the variability in the CP group (top curve) is visibly larger than that of the TD group.

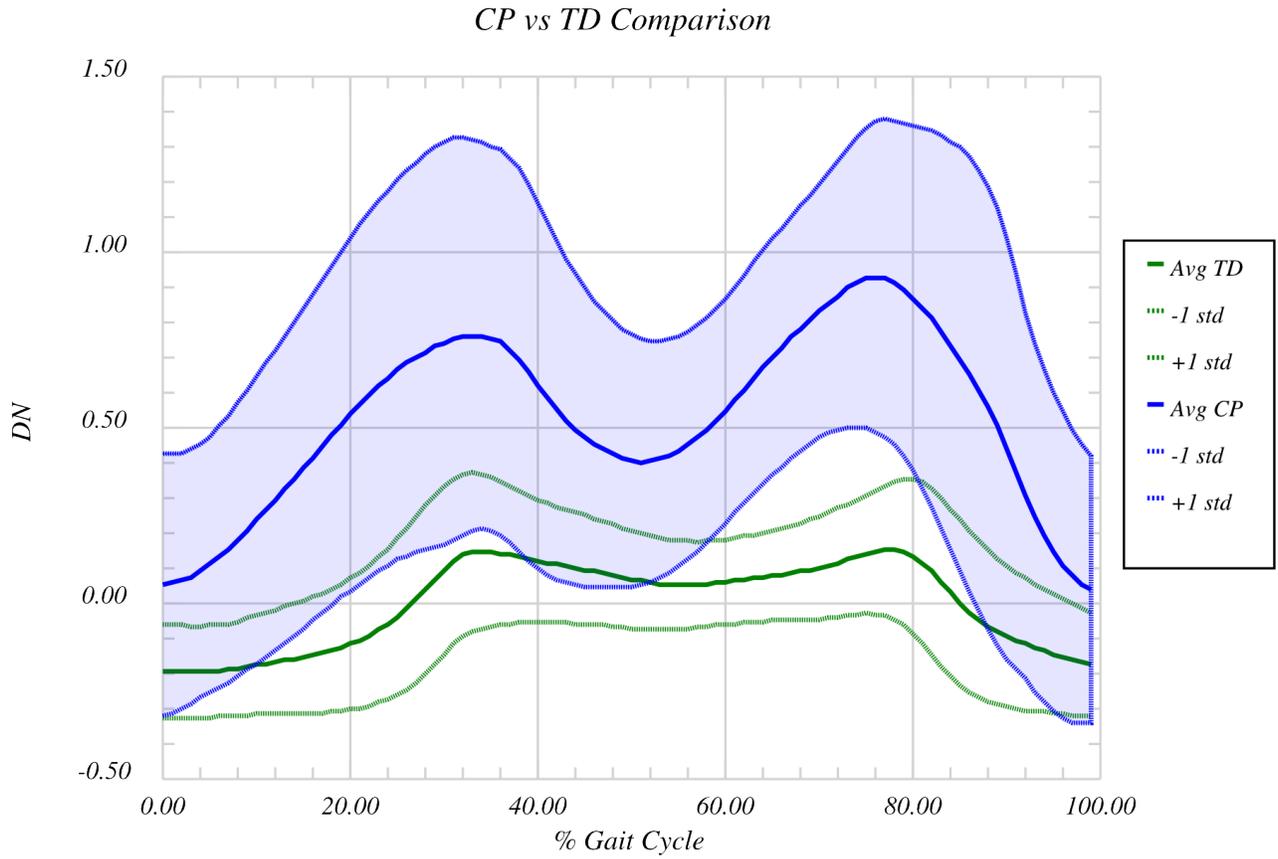